\documentclass[appendixfloats]{emulateapj}

\shortauthors{Auger et al.}
\shorttitle{Dark Matter Contraction and Light IMFs}

\newcommand{\be}{\begin{equation}}
\newcommand{\ee}{\end{equation}}
\newcommand{\bea}{\begin{eqnarray}}
\newcommand{\eea}{\end{eqnarray}}
\newcommand{\lnp}{$\langle{\rm ln P}\rangle$}

\begin{document}

\title{Dark Matter Contraction and the Stellar Content of Massive Early-type Galaxies: Disfavoring ``Light'' Initial Mass Functions}

\author{M.~W.~Auger\altaffilmark{1,$\dagger$}, T.~Treu\altaffilmark{1,2},  R. Gavazzi\altaffilmark{3}, A. S. Bolton\altaffilmark{4}, L. V. E. Koopmans\altaffilmark{5}, P. J. Marshall\altaffilmark{6}}
\altaffiltext{1}{Department of Physics, University of California, Santa Barbara, CA 93106, USA}
\altaffiltext{2}{Packard Fellow}
\altaffiltext{3}{Institut d'Astrophysique de Paris, UMR7095 CNRS \& Univ. Pierre et Marie Curie, 98bis Bvd Arago, F-75014 Paris, France}
\altaffiltext{4}{Department of Physics and Astronomy, University of Utah, Salt Lake City, UT 84112}
\altaffiltext{5}{Kapteyn Astronomical Institute, University of Groningen, P.O. Box 800, 9700AV Groningen, The Netherlands}
\altaffiltext{6}{Kavli Institute for Particle Astrophysics and Cosmology, Stanford University, Stanford, CA 94305, USA}
\altaffiltext{$\dagger$} {\texttt mauger@physics.ucsb.edu}

\begin{abstract}
We use stellar dynamics, strong lensing, stellar population synthesis
models, and weak lensing shear measurements to constrain the dark
matter (DM) profile and stellar mass in a sample of 53 massive
early-type galaxies. We explore three DM halo models (unperturbed
Navarro Frenk \& White [NFW] halos and the adiabatic contraction
models of Blumenthal and Gnedin) and impose a model for the relationship between the stellar
and virial mass (i.e., a relationship for the star-formation
efficiency as a function of halo mass). We show that, given our model
assumptions, the data clearly prefer a Salpeter-like initial mass
function (IMF) over a lighter IMF (e.g., Chabrier or Kroupa),
irrespective of the choice of DM halo. In addition, we find that the
data prefer at most a moderate amount of adiabatic contraction (Blumenthal adiabatic contraction is strongly disfavored) and are
only consistent with no adiabatic contraction (i.e., a NFW halo) if a
mass-dependent IMF is assumed, in the sense of a more massive
normalization of the IMF for more massive halos.
\end{abstract}

\keywords{
   galaxies: elliptical and lenticular, cD --- galaxies: halos --- galaxies: structure --- dark matter
}

\section{INTRODUCTION}
Numerical simulations of cold dark matter (CDM) halos predict density
distributions with an inner logarithmic slope $d{\rm ln}\rho/d{\rm
ln}r = -1$ and a slope of $d{\rm ln}\rho/d{\rm ln}r = -3$ at large
radii \citep[][hereafter NFW]{nfw,navarro}. However, the inclusion of
baryons -- and the dissipational processes associated with them -- in
galaxy formation simulations may significantly alter the central dark
matter distribution. The DM may show an increased central
density due to radiative dissipation from the baryons that depends on
the mass of stars formed in the galaxy \citep[e.g., the adiabatic
contraction, or AC, models of][hereafter B86 and
G04]{blumenthal,gnedin} or dynamical friction may counter-balance this
effect \citep{abadi} and could result in cored profiles
\citep[e.g.,][]{elzant,nipoti}. The form of the CDM halo is critical
for understanding the relative importance of baryonic and total mass
in governing the structure of early-type galaxies (ETGs).

Strong lensing can measure masses to a few percent
precision and is therefore a particularly powerful probe of
this regime, by itself and in combination with other techniques
\citep[][and references therein]{TTreview}. Studies have previously
used strong and weak gravitational lensing (SL and WL) to constrain the
distribution of DM for massive ETGs and found that the data are
broadly and on average consistent with a NFW halo
\citep[e.g.,][]{slacsiv}. However, the SL and WL
analysis alone cannot simultaneously probe the stellar mass M$_*$ and
central dark matter slope due to degeneracies
\citep{lagattuta}; additional information must be included to
disentangle the contributions of stars and CDM to the central
density distribution (similar to the disk-halo degeneracy for
spiral galaxies). \citet{schulz} have recently used WL and
stellar velocity dispersions to investigate evidence for AC in SDSS
ETGs. They find that the stellar velocity dispersions require more
central mass than would be inferred from the stellar mass (assuming a
Kroupa initial mass function, or IMF) and the central DM mass
extrapolated from a NFW halo fit to the WL data. However,
their findings are also consistent with no AC if a Salpeter IMF is
assumed for the stellar component.

\citet{jiang} point out that the degeneracies between the central 
CDM slope and M$_*$ can only be broken by using at least three mass
probes. They use SL and dynamics to probe the central
baryonic and CDM distribution of ETGs and include an ensemble
measurement of the halo mass from WL data to find that AC
models are favored over a NFW halo. However, \citet{jiang} assume a
redshift-dependent but otherwise constant stellar mass-to-light ratio
M$_*$/L for each of the galaxies, although the IMF (and therefore
M$_*$/L) may be non-universal \citep{t10}. A full understanding of the
central mass distribution of galaxies requires constraints on both the
stellar and CDM components as a function of mass. Although this joint
analysis is more complex and there are residual
degeneracies, it provides a rare opportunity to constrain directly the
IMF of distant massive ETGs where resolved stellar population
diagnostics are not available \citep[see, e.g.,][and references
therein]{cappellari09,bastian}.

In this Letter we present simultaneous constraints on M$_*$ and the
form of the central dark matter distribution for a sample of massive
ETGs. We use SL, stellar dynamics, stellar population
synthesis (SPS) stellar mass estimates M$_{*}^{\rm SPS}$ given an IMF,
and WL shear measurements to constrain a model for the
relationship between stellar and virial mass. We allow for several different DM halo profiles and we fit for an IMF mismatch parameter
relating M$_*$ to M$_*^{\rm SPS}$ \citep{t10}.  The two main
improvements over the analysis by \citet{t10} are the inclusion of
WL data and the adoption of more flexible models. We use a standard flat $\Lambda$CDM cosmology with $\Omega_{\rm M} = 0.3$ and $h=0.7$.

\section{The Data and Model}

Our sample is derived from the Sloan Lens ACS Survey
\citep[SLACS;][]{slacsi,slacsiv,slacsv,slacsix}, which consists
of nearly 100 strong gravitational lensing galaxies. We exclude 6
galaxies which appear to be structurally different than the rest of
the SLACS early-type lenses \citep{slacsx}.  These systems are
significant outliers from the otherwise very tight relation between
effective radius, velocity dispersion, total central mass, and stellar
mass; this is likely due to systematic errors in the stellar velocity
dispersions of these systems \citep[also see][]{jiang}.
We therefore use a subsample of 53 ETGs that have well-determined central stellar velocity dispersions, SPS stellar masses assuming \citet{chabrier} and
\citet{salpeter} IMFs, and central projected mass estimates from
SL. We also use WL shear measurements for a subset of 44 lenses with deep ACS imaging. The details of how these data are derived can be found elsewhere
\citep{slacsiv,slacsv,slacsix,slacsx}. We emphasize
that the SLACS lens galaxies are indistinguishable from twin ETGs
selected from SDSS to have the same velocity dispersion and redshift
and can therefore be considered as representative of the overall
population of massive ETGs \citep{slacsii,slacsviii,slacsix,slacsx}.

We model the lenses as two-component mass distributions consisting of
stars and a CDM halo. Both components are modeled as spherical since
our data do not constrain the halo flattening and our dynamical
analysis is restricted to spherical Jeans modeling. Comparisons with
non-spherical models shows that the most relevant quantities such as
mass-density slope and total mass are relatively insensitive to this
approximation \citep[e.g.,][]{barnabe}.

The stars are modeled with the \citet{hernquist} profile and the scale
radius is set by the rest-frame $V$-band effective radius, $r_a =
0.551 r_{\rm e}$; our conclusions are unchanged if we adopt a \citet{jaffe} profile.
We assume isotropic orbits, consistent with typical findings for massive
ETGs \citep{barnabe}. Mild radial anisotropy
\citep{k09} only alters the inferred velocity dispersions by a few percent, i.e. an amount smaller than the errors, and therefore does not change our conclusions.

We investigate three different density profiles for the CDM halos. First,
we consider the NFW profile with the concentration fixed by $c({\rm M_{\rm vir}}) \propto {\rm M}_{\rm vir}^{-0.094}$ following \citet{maccio},
neglecting intrinsic scatter for computational simplicity. We also modify the NFW halo
to include AC using the prescriptions of B86 and
G04, with the baryon fraction set by Equation 1 and the scale radius given by $r_a$; these three profiles are the most widely adopted descriptions for CDM halos \citep[e.g.,][]{jiang,schulz,napolitano}, and the NFW (B86) halo has the lowest (highest) central DM density.

We fix the relationship between stellar and virial mass to have the
form suggested by \citet{moster} (who use a Kroupa IMF) based upon abundance matching techniques.
We assume
\be
{\rm M}_* \equiv {\rm M}_*^{\rm SPS} = {\rm M}_{*,0} \frac{\left({\rm M}_{\rm vir}/{\rm M}_{\rm vir,0}\right)^{\gamma_1}}{\left[1 + \left({\rm M}_{\rm vir}/{\rm M}_{\rm vir,0}\right)^\beta\right]^{(\gamma_1-\gamma_2)/\beta}}
\ee
where M$_*^{\rm SPS}$ is the stellar mass inferred from the SPS models assuming a Salpeter IMF, $\gamma_1 = 7.17$, $\gamma_2 =  0.201$, and $\beta =  0.557$
and we fit for M$_{*,0}$ and M$_{\rm vir,0}$. We note that our results
are qualitatively the same if we use different values for $\gamma_1,
\gamma_2,$ and $\beta$ \citep[e.g., from][]{dutton}, although the goodness-of-fit changes; we will explicitly fit for these parameters in a future paper to ascertain the extent to which one model prefers a given set of parameters over another.

We also allow for an offset between the model stellar mass and the SPS
stellar mass, equivalent to a mass-dependent IMF (referred to as the `Free' IMF model). The offset has the form
\be
\label{E_imf}
{\rm log}\frac{\rm M_*^{\rm SPS}}{\rm M_*} = -\eta {\rm log}\frac{\rm
M_*}{10^{11} {\rm M}_\odot} - \log \alpha,
\ee 
where $\alpha$ is equivalent to that introduced by
\citet{t10} and $\eta$ is equivalent to the slope of Equation 5 in \citet{t10}. We adopt the M$_*^{\rm SPS}$ derived by
\citet{slacsix}, although our results are robust with respect to
changes in the SPS models and choice of priors \citep{t10} due to the relative simplicity of the older stellar populations of the SLACS ETGs. Uncertainties in SPS models of evolved galaxies \citep{conroy,maraston} may lead to systematic offsets in M$_*^{\rm SPS}$ of order $\sim 0.05$ dex \citep[e.g.,][]{t10}.

We perform a Markov Chain Monte Carlo simulation to fit the models to the data; we propose M$_{\rm vir}$ for each lens and use Equation 1 (and 2 for the Free IMF model) to predict a stellar mass. These masses normalize the halo and bulge, and we use these profiles to predict the stellar kinematics, SL, and WL signals for each galaxy. We compare these quantities and the model stellar masses with the observed quantities on a galaxy-by-galaxy basis (i.e., we do not bin or average measurements, including the WL) to optimize the free model parameters and determine goodness-of-fit values.

\section{Results}
\begin{figure*}[!ht]
\begin{center}
 \includegraphics[width=0.98\textwidth,clip]{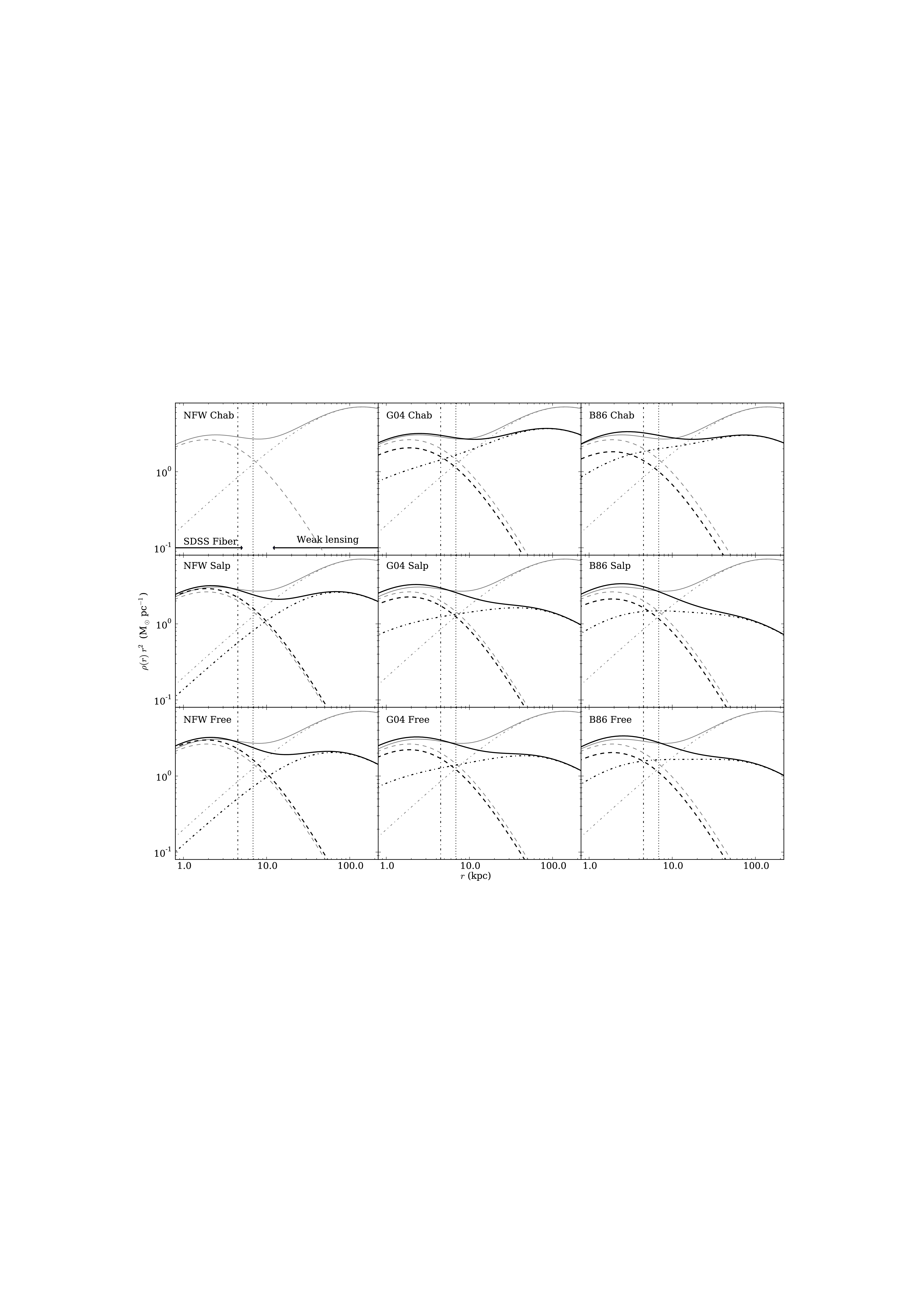}
\end{center}
 \vspace{-2 mm}
 \caption{Bulge (dashed), halo (dot-dashed), and total mass (solid) profiles for each of the models investigated. The fiducial model, a NFW halo with Chabrier IMF, is shown in grey in each panel for comparison, and the vertical lines indicate the effective radius (dotted) and Einstein radius (dot-dashed); the radii probed by the stellar velocity dispersion and WL data are illustrated in the upper left panel. 
Note that the total mass profiles are approximately isothermal within the effective radius \citep[e.g.,][]{slacsiii,k09,slacsx}. The Chabrier IMF models require large halo masses in order to provide enough projected mass within the Einstein radius to fit the SL data.}
 \label{F_bulgehalo}
\end{figure*}

\begin{figure}[!ht]
\begin{center}
 \includegraphics[width=0.48\textwidth,clip]{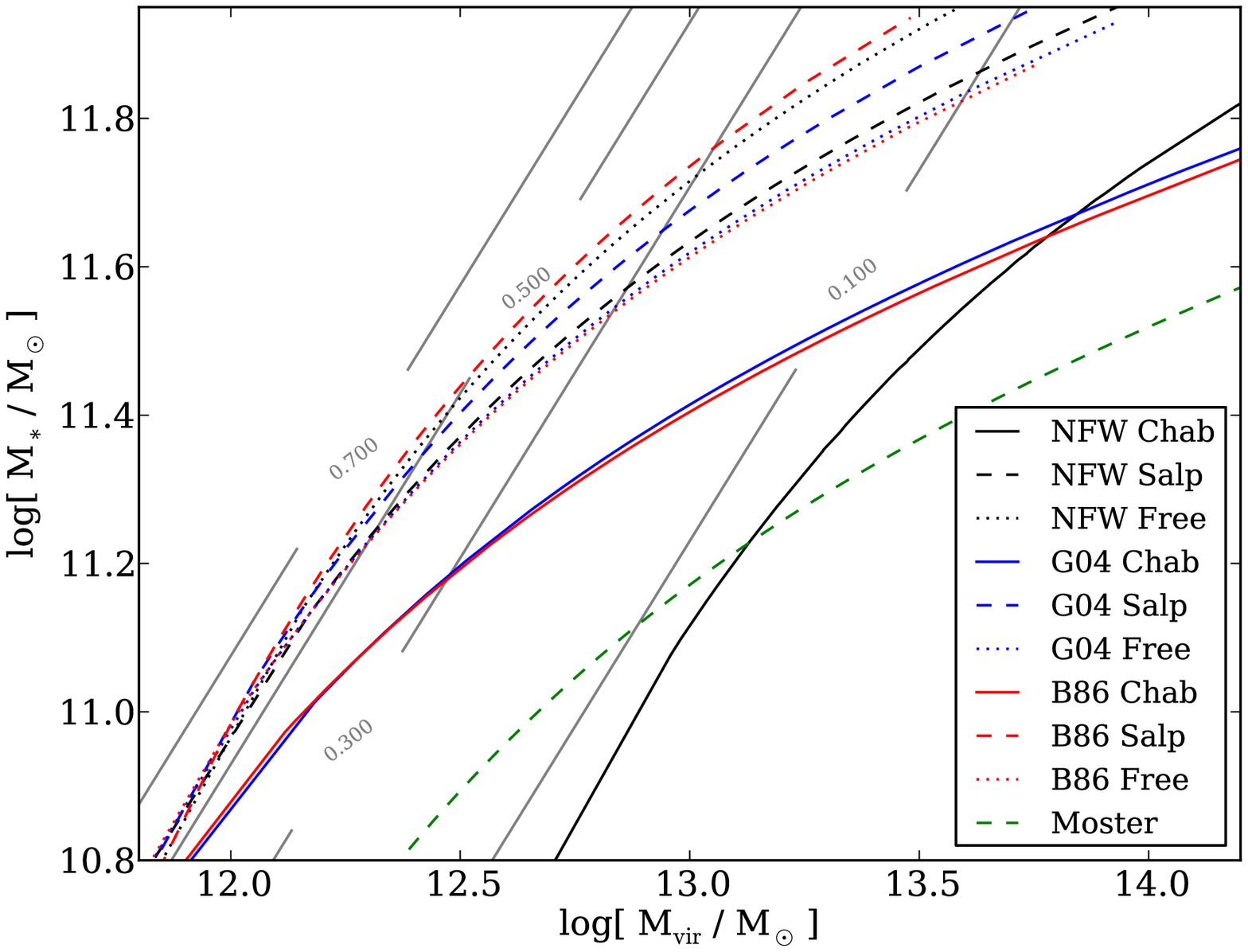}
\end{center}
 \caption{Best-fit M$_{\rm vir}$-M$_*$ relation for each of the bulge$+$halo models considered. Note that the models with a Chabrier IMF (solid lines) require substantially larger halo masses at fixed stellar mass in order to fit the SL data. The grey lines indicate contours of constant star-formation efficiency.}
 \label{F_mstar_mvir}
\end{figure}

The best-fit parameters and $\chi^2$ for each model are listed in
Table \ref{T_results}, and we show the best-fit mass distributions for
a characteristic lens system ($L_V = 10^{10.85} L_\odot$) in Figure
\ref{F_bulgehalo}. This illustrates that the low normalization
of the Chabrier IMF requires a more massive CDM halo to
increase the central projected mass and be consistent with the mass
required by the SL data (this is true for the NFW
\emph{and} AC models). However, increasing the halo mass also increases the
scale radius because the virial radius increases and the concentration
becomes smaller \citep[e.g.,][]{maccio}; therefore, the halo mass
grows at a faster rate than the projected dark matter mass inside the
Einstein radius. The net effect is that very massive halos are needed
to match the SL data if a Chabrier IMF is used (see Figure \ref{F_mstar_mvir}).

These massive halos cannot fit the SL masses \emph{and}
the WL shears (the inferred M$_*$ values are also much higher than the measurements), and the data therefore clearly
disfavor a Chabrier IMF (Table \ref{T_results}), consistent with the
results from SL and dynamical constraints only
\citep{t10}. The G04 model is slightly preferred over the NFW halo
for the Salpeter IMF while the B86 model is disfavored. The
normalization and break point of the M$_{\rm vir}$-M$_*$ relation are
consistent between the NFW and G04 model with a Salpeter IMF, and both
are also consistent with the results from \citet{moster} when the
$\sim 0.2$ dex offset between Kroupa and Salpeter IMFs is
taken into account.

Figure \ref{F_mstar_mvir} illustrates that the favored models require
significantly higher star-formation efficiencies compared to the values of $\approx 0.02 - 0.1$ (for a Chabrier IMF; double this for Salpeter) found for massive ETGs from galaxy-galaxy WL studies \citep[e.g.,][]{mandelbaum}, which may point to a generic inadequacy of the family of models considered here. Interestingly, simulations of galaxy formation may require similarly high star-formation efficiencies to reproduce the
observed isothermal central mass distributions of massive ETGs \citep{duffy} and suggest that the high efficiency may not be due to our simplifying assumptions.

\begin{deluxetable*}{llcccccc}
\tabletypesize{\scriptsize}
\tablecolumns{8}
\tablewidth{0pc}
\tablecaption{Best Fitting Parameters and Goodness of Fits}
\tablehead{
 \colhead{Halo} &
 \colhead{IMF}  &
 \colhead{log[ M$_{*,0}$ / M$_\odot$ ]} &
 \colhead{log[ M$_{\rm vir,0}$ / M$_\odot$ ]} &
 \colhead{$\eta$} &
 \colhead{log $\alpha$} &
 \colhead{$\chi^2$/d.o.f.} &
 \colhead{\lnp}
}
\startdata
NFW  &  Chab  &  $11.38\pm 0.09$  &  $11.32\pm 0.24$  &  \nodata  &  \nodata  &  2.68  &  -19.16 \\
  &  Salp  &  $11.29\pm 0.04$  &  $10.38\pm 0.06$  &  \nodata  &  \nodata  &  1.18  &  266.54 \\
  &  Free  &  $11.44\pm 0.06$  &  $10.52\pm 0.07$  &  $0.08\pm0.04$  &  $0.03\pm0.03$  &  1.04  &  284.16 \\
G04  &  Chab  &  $10.97\pm 0.04$  &  $10.11\pm 0.11$  &  \nodata  &  \nodata  &  1.66  &  189.76 \\
  &  Salp  &  $11.36\pm 0.05$  &  $10.43\pm 0.05$  &  \nodata  &  \nodata  &  1.12  &  268.98 \\
  &  Free  &  $11.26\pm 0.06$  &  $10.34\pm 0.06$  &  $0.03\pm0.05$  &  $-0.08\pm0.02$  &  1.08  &  281.58 \\
B86  &  Chab  &  $10.94\pm 0.04$  &  $10.07\pm 0.10$  &  \nodata  &  \nodata  &  1.42  &  228.43 \\
  &  Salp  &  $11.47\pm 0.06$  &  $10.54\pm 0.06$  &  \nodata  &  \nodata  &  1.26  &  234.00 \\
  &  Free  &  $11.25\pm 0.06$  &  $10.33\pm 0.07$  &  $0.01\pm0.05$  &  $-0.11\pm0.02$  &  1.13  &  269.05 \\
\enddata
\label{T_results}
\tablecomments{The `Free' IMF is from Equation \ref{E_imf}; for $\eta = 0$, log $\alpha$ = 0 for a Salpeter IMF and log~$\alpha~\approx~-0.25$ (log $\alpha \approx -0.2$) for a Chabrier (Kroupa) IMF. The $\chi^2$ is given for a model with the mean parameters listed in the table, although the interpretation of $\Delta \chi^2$ is difficult due to the large number of degrees of freedom; \lnp is the mean of the natural log of the posterior probability. It is also difficult to directly interpret $\Delta \langle{\rm ln P}\rangle$, but values of several tens clearly indicate a strong preference for one model over another.}
\end{deluxetable*}

\begin{figure}[!ht]
\begin{center}
 \includegraphics[width=0.48\textwidth,clip]{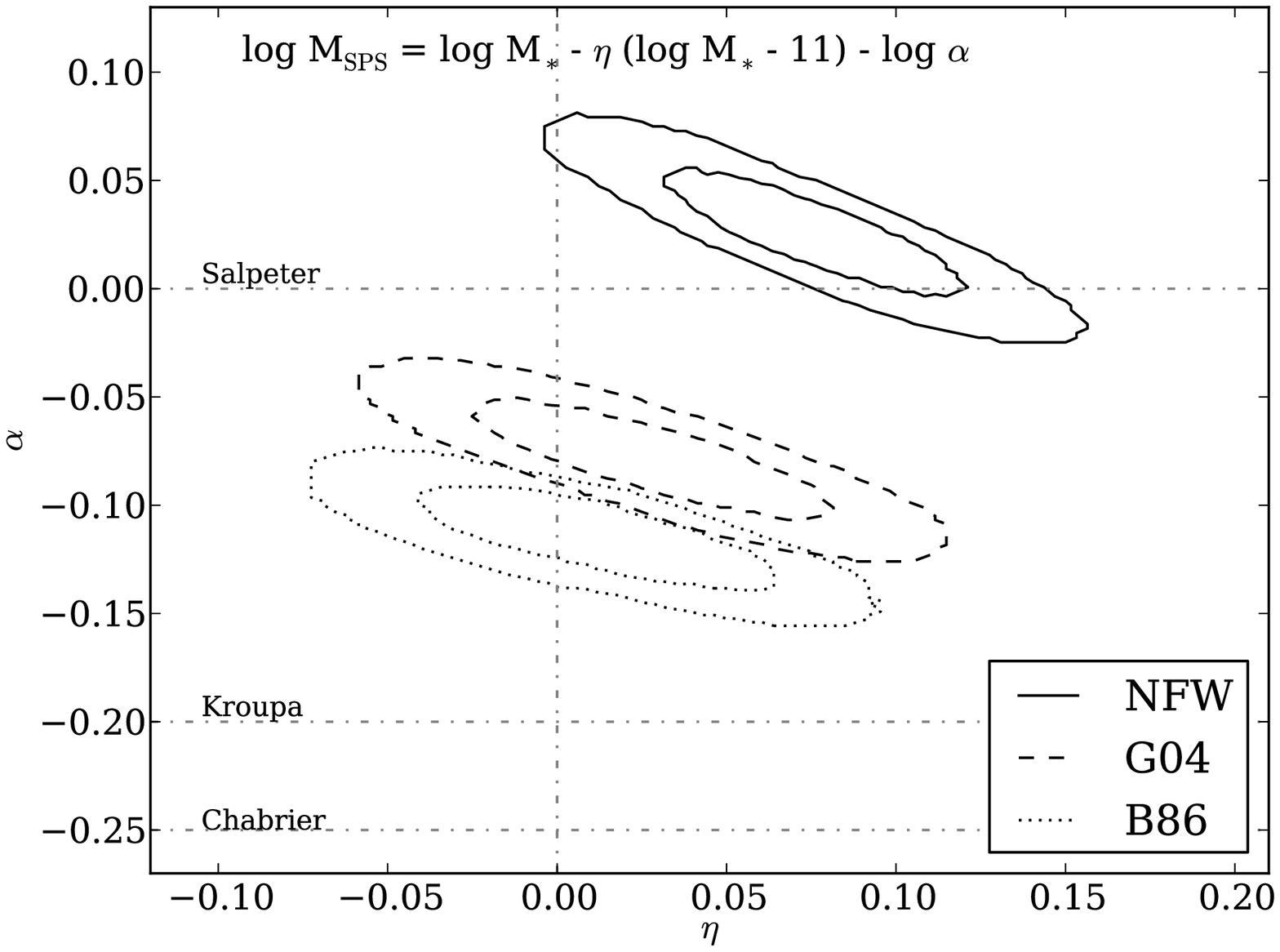}
\end{center}
 \caption{Constraints on the M$_{\rm SPS}$ mismatch model (Equation \ref{E_imf}) assuming a NFW halo (solid contours), G04 halo (dashed contours), or B86 halo (dotted contours); the inner (outer) contour encloses 68\% (95\%) of the probability. The SPS masses were determined assuming the same IMF (Salpeter) for each galaxy and a non-zero $\eta$ may therefore signal a mass-dependent IMF. The vertical line indicates a non-evolving IMF, and the horizontal lines denote the expected $\alpha$ for common IMFs, as indicated.}
 \label{F_eta_alpha}
\end{figure}

The results for the free IMF model are somewhat more difficult to
interpret due to the addition of two extra parameters; the fit is
significantly improved compared to the fixed IMF models, as is
expected when allowing for a more flexible model. The NFW model
and G04 model fit the data equally well, but again the B86 model is
disfavored. Note, however, that the form of the M$_{\rm vir}$-M$_*$
relation that we are using was derived assuming a constant IMF;
allowing for a varying IMF would necessarily change the structural
parameters $\gamma_1, \gamma_2,$ and $\beta$. Nevertheless, the
present model indicates that even the AC halos require the IMF to have a
slight trend with mass in the sense that more massive galaxies require
more Salpeter-like IMFs. The trend is stronger with the NFW halo
(Figure \ref{F_eta_alpha}) and is consistent with the relation found
by \citet{t10}.

\section{Discussion}

There is a growing consensus that the fraction of DM at the
center of massive ETGs increases with galaxy mass
\citep{tortora,napolitano,graves,slacsx}. This increase could be due
to a genuine increase in cold dark matter, perhaps as the result of AC
mediated by the detailed accretion history of halos
\citep[e.g.,][]{abadi,lackner}. Alternatively, the trends may be the
result of a mass-dependent IMF leading to more baryonic DM
(in the form of low-mass stars or stellar remnants, depending on
whether the IMF is bottom-heavy or top-heavy, respectively) in more
massive galaxies \citep[e.g.,][]{t10}, perhaps due to cosmic evolution
of the IMF \citep{vandokkum} and a mass-dependent formation redshift
for the stellar mass \citep[e.g.,][]{thomas,vanderwel}.

\subsection{Ruling Out ``Light'' IMFs for Massive ETGs}

We find that our sample of massive ETGs is inconsistent with a
Chabrier IMF, even after allowing for strong CDM halo
contraction and a mass-dependent IMF. Indeed, our data disfavor common
``light'' IMFs \emph{generally}; the stellar mass derived assuming a
\citet{kroupa} IMF is only $\sim 0.06$ dex greater than M$_*$ from a
Chabrier IMF, which
is not able to account for the $\alpha$ values that we find (Table
\ref{T_results}). We note that based on our data alone we cannot
distinguish whether this is due to a higher abundance of low-mass
stars (Salpeter-like), or of higher-mass stars and their neutron star
and black hole remnants. 

These results may appear to be inconsistent with those of
\citet{cappellari}, who find that an IMF with a higher
normalization than Kroupa (e.g., Salpeter) leads to stellar masses
that are sometimes greater than the dynamical masses. However, we note
that the tension is entirely due to fast-rotating or lower-mass
galaxies, neither of which are typical of the SLACS lenses
\citep[e.g.,][]{barnabe,slacsx}; the SAURON data are consistent with
our finding that massive ETGs do not have bottom-light IMFs \citep{cappellari}.

\subsection{Adiabatic Contraction and a Varying IMF}

The data also are opposed to strong adiabatic contraction, as
modeled by B86. However, we cannot currently
distinguish between a halo that has undergone mild AC and a NFW
halo. This is partially due to imposing the form of the M$_{\rm
vir}$-M$_*$ relation; if we use the \citet{dutton} parameters for Equation 1
we find that for a fixed IMF the G04 model is strongly
favored over the NFW model (and B86 is still disfavored), although we
are again unable to distinguish between the two halos when the IMF is
allowed to vary with mass. 

\citet{schulz} similarly found that a Kroupa IMF and adiabatically
contracted halo or Salpeter IMF and no AC adequately fit the
data, but they do not consider a mass-dependent IMF.
Likewise, \citet{jiang} find that a Salpeter-like IMF and AC model is preferred 
\citep[they do not directly compare their
stellar masses with SPS model stellar masses, but their derived
$B$-band M$_*$/L = 7.2 M$_\odot$/L$_\odot$ is consistent with the
M$_*$/L we find for the SLACS lenses assuming a Salpeter IMF,
e.g.,][]{slacsx}.  Note, however, that they do not test for a mass-dependent IMF (their M$_*$/L is independent of mass).

\subsection{Beyond Structural Constraints}

The structural decomposition into bulge+halo models is useful but may
not be sufficient to fully understand how the IMF varies as a function
of galaxy properties. We conclude by considering how non-structural
constraints might be used to further discriminate between CDM and
a non-universal IMF.

\citet{napolitano} use stellar dynamics and SPS models to determine
the DM fraction for a sample of ETGs and compare these with
the predicted DM fractions from a suite of bulge+halo toy
models. They find that a Kroupa IMF with AC or Salpeter IMF with no
AC best represent their data, although they do not directly fit the
models and do not allow for a mass-dependent star-formation
efficiency. Nevertheless, they find an intriguing trend in which the
IMF does not depend on mass but may depend strongly on age in the
sense that older galaxies have more Salpeter-like IMFs. As noted by
\citet{napolitano}, this is the \emph{opposite} trend than is expected
from our relations.
The cause of this discrepancy is unclear but may be related to how the samples are chosen. The SLACS lenses are at higher redshifts than the \citet{napolitano} galaxies and are selected to have spectra that match old stellar templates; they would therefore only span ages between approximately 5 and 10 Gyr, for which the age-IMF trend is essentially flat.

Likewise, \citet{graves} use trends between mass and metallicity
([Fe/H], [Mg/H], and [Mg/Fe]) to investigate the role of the IMF on
the central dark matter fraction. They show that the
$\alpha$-enhancement of galaxies (e.g., the [Mg/Fe] ratio) is
mass-dependent and could therefore signal a more top-heavy IMF in more
massive galaxies such that the relative number of core-collapse
supernovae increases in more massive galaxies. However, invoking more
supernovae yields to super-abundant metallicities compared to the
data, and another mechanism must therefore be posited in order to
remove these metals \citep{graves}. 

\section{Conclusions}
We have demonstrated that, given our standard assumptions, the IMF of
massive ETGs cannot be ``light'', even in the presence of significant
adiabatic contraction. Furthermore, the IMF is only consistent with
being universal if the central dark matter profile is mass dependent,
as in the AC  models presented here.  It is clear that
if one wants to preserve a light and universal IMF one has to abandon
standard assumptions about dark matter halos, like the NFW profile.
However, there are two important caveats. First, better constraints on
the star-formation efficiency must be obtained from the data in order
to draw definitive conclusions about the role of a mass-dependent IMF
relative to CDM halo contraction. Second, although our conclusions are
robust with respect to a wide variety of changes in our assumptions,
we have not exhausted {\it all} possible families of theoretical
models.

We will address these caveats, quantify the relationship between the central
dark matter profile and the IMF in more detail, and quantify the consequences of our various assumptions in a forthcoming paper \citep[][in preparation]{haloimf}.

\acknowledgments
We thank the referee for his/her helpful comments, and Aaron Dutton and Matteo Barnab{\`e} for their suggestions. We acknowledge support from the NSF through CAREER award NSF-0642621, from the Packard Foundation through a Packard Fellowship to TT, and from NASA, through HST grants 11202, 10798, and 10494. RG is supported by the Centre National des Etudes Spatiales.


\begin{thebibliography}{}

\bibitem[Abadi et al.(2009)]{abadi} Abadi, M.~G., Navarro, J.~F., Fardal, M., Babul, A., \& Steinmetz, M.\ 2009, arXiv:0902.2477 

\bibitem[Auger et al.(2009)]{slacsix} Auger, M.~W., Treu, T., Bolton, A.~S., Gavazzi, R., Koopmans, L.~V.~E., Marshall, P.~J., Bundy, K., \& Moustakas, L.~A.\ 2009, \apj, 705, 1099 

\bibitem[Auger et al.(2010a)]{slacsx} Auger, M.~W., et al.\ 2010a, submitted to ApJ

\bibitem[Auger et al.(2010b)]{haloimf} Auger, M.~W., et al.\ 2010b, in preparation

\bibitem[Barnab{\`e} et al.(2009)]{barnabe} Barnab{\`e}, M., Czoske, O., Koopmans, L.~V.~E., Treu, T., Bolton, A.~S., \& Gavazzi, R.\ 2009, \mnras, 399, 21 

\bibitem[Bastian et al.(2010)]{bastian} Bastian, N., Covey, K.~R., \& Meyer, M.~R.\ 2010, arXiv:1001.2965

\bibitem[Blumenthal et al.(1986)]{blumenthal} Blumenthal, G.~R., Faber, S.~M., Flores, R., \& Primack, J.~R.\ 1986, \apj, 301, 27 

\bibitem[Bolton et al.(2006)]{slacsi} Bolton, A.~S., Burles, S., Koopmans, L.~V.~E., Treu, T., \& Moustakas, L.~A.\ 2006, \apj, 638, 703

\bibitem[Bolton et al.(2008)]{slacsv} Bolton, A.~S., Burles, S., Koopmans, L.~V.~E., Treu, T., Gavazzi, R., Moustakas, L.~A., Wayth, R., \& Schlegel, D.~J.\ 2008, \apj, 682, 964

\bibitem[Cappellari et al.(2006)]{cappellari} Cappellari, M., et al.\ 2006, \mnras, 366, 1126 

\bibitem[Cappellari et al.(2009)]{cappellari09} Cappellari, M., et al.\ 2009, arXiv:0908.1904 

\bibitem[Chabrier(2003)]{chabrier} Chabrier, G.\ 2003, \pasp, 115, 763 

\bibitem[Conroy \& Gunn(2010)]{conroy} Conroy, C., \& Gunn, J.~E.\ 2010, \apj, 712, 833

\bibitem[Duffy et al.(2010)]{duffy} Duffy, A.~R., Schaye, J., Kay, S.~T., Vecchia, C.~D., Battye, R.~A., \& Booth, C.~M.\ 2010, \mnras, 624 

\bibitem[Dutton et al.(2010)]{dutton} Dutton, A.~A., Conroy, C., van den Bosch, F.~C., Prada, F., \& More, S.\ 2010, arXiv:1004.4626 

\bibitem[El-Zant et al.(2004)]{elzant} El-Zant, A.~A., Hoffman, Y., Primack, J., Combes, F., \& Shlosman, I.\ 2004, \apjl, 607, L75 

\bibitem[Gavazzi et al.(2007)]{slacsiv} Gavazzi, R., Treu, T., Rhodes, J.~D., Koopmans, L.~V.~E., Bolton, A.~S., Burles, S., Massey, R.~J., \& Moustakas, L.~A.\ 2007, \apj, 667, 176 

\bibitem[Gnedin et al.(2004)]{gnedin} Gnedin, O.~Y., Kravtsov, A.~V., Klypin, A.~A., \& Nagai, D.\ 2004, \apj, 616, 16 

\bibitem[Graves \& Faber(2010)]{graves} Graves, G.~J., \& Faber, S.~M.\ 2010, arXiv:1005.0014 

\bibitem[Hernquist(1990)]{hernquist} Hernquist, L.\ 1990, \apj, 356, 359

 \bibitem[Jaffe(1983)]{jaffe} Jaffe, W.\ 1983, \mnras, 202, 995 

\bibitem[Jiang \& Kochanek(2007)]{jiang} Jiang, G., \& Kochanek, C.~S.\ 2007, \apj, 671, 1568 

\bibitem[Koopmans et al.(2006)]{slacsiii} Koopmans, L.~V.~E., Treu, T., Bolton, A.~S., Burles, S., \& Moustakas, L.~A.\ 2006, \apj, 649, 599

\bibitem[Koopmans et al.(2009)]{k09} Koopmans, L.~V.~E., et al.\ 2009, \apjl, 703, L51 

\bibitem[Kroupa(2001)]{kroupa} Kroupa, P.\ 2001, \mnras, 322, 231

\bibitem[Lackner \& Ostriker(2010)]{lackner} Lackner, C.~N., \& Ostriker, J.~P.\ 2010, \apj, 712, 88 


\bibitem[Lagattuta et al.(2009)]{lagattuta} Lagattuta, D.~J., et al.\ 2009, arXiv:0911.2236 

\bibitem[Macci{\`o} et al.(2008)]{maccio} Macci{\`o}, A.~V., Dutton, A.~A., \& van den Bosch, F.~C.\ 2008, \mnras, 391, 1940 

\bibitem[Mandelbaum et al.(2006)]{mandelbaum} Mandelbaum, R., Seljak, U., Kauffmann, G., Hirata, C.~M., \& Brinkmann, J.\ 2006, \mnras, 368, 715 

\bibitem[Maraston et al.(2009)]{maraston} Maraston, C., Str{\"o}mb{\"a}ck, G., Thomas, D., Wake, D.~A., \& Nichol, R.~C.\ 2009, \mnras, 394, L107

\bibitem[Moster et al.(2010)]{moster} Moster, B.~P., Somerville, R.~S., Maulbetsch, C., van den Bosch, F.~C., Macci{\`o}, A.~V., Naab, T., \& Oser, L.\ 2010, \apj, 710, 903 

\bibitem[Napolitano et al.(2010)]{napolitano} Napolitano, N.~R., Romanowsky, A.~J., \& Tortora, C.\ 2010, \mnras, 793 

\bibitem[Navarro et al.(1996)]{nfw} Navarro, J.~F., Frenk, C.~S., \& White, S.~D.~M.\ 1996, \apj, 462, 563 

\bibitem[Navarro et al.(2010)]{navarro} Navarro, J.~F., et al.\ 2010, \mnras, 402, 21 

\bibitem[Nipoti et al.(2004)]{nipoti} Nipoti, C., Treu, T., Ciotti, L., \& Stiavelli, M.\ 2004, \mnras, 355, 1119 

\bibitem[Salpeter(1955)]{salpeter} Salpeter, E.~E.\ 1955, \apj, 121, 161 

\bibitem[Schulz et al.(2009)]{schulz} Schulz, A.~E., Mandelbaum, R., \& Padmanabhan, N.\ 2009, arXiv:0911.2260 

\bibitem[Thomas et al.(2005)]{thomas} Thomas, D., Maraston, C., Bender, R., \& Mendes de Oliveira, C.\ 2005, \apj, 621, 673

\bibitem[Tortora et al.(2009)]{tortora} Tortora, C., Napolitano, N.~R., Romanowsky, A.~J., Capaccioli, M., \& Covone, G.\ 2009, \mnras, 396, 1132 

\bibitem[Treu et al.(2006)]{slacsii} Treu, T., Koopmans, L.~V., Bolton, A.~S., Burles, S., \& Moustakas, L.~A.\ 2006, \apj, 640, 662

\bibitem[Treu et al.(2009)]{slacsviii} Treu, T., Gavazzi, R., Gorecki, A., Marshall, P.~J., Koopmans, L.~V.~E., Bolton, A.~S., Moustakas, L.~A., \& Burles, S.\ 2009, \apj, 690, 670

\bibitem[Treu et al.(2010)]{t10} Treu, T., Auger, M.~W., Koopmans, L.~V.~E., Gavazzi, R., Marshall, P.~J., \& Bolton, A.~S.\ 2010, \apj, 709, 1195 

\bibitem[Treu(2010)]{TTreview} Treu, T.\ 2010, arXiv:1003.5567

\bibitem[van der Wel et al.(2009)]{vanderwel} van der Wel, A., Bell, E.~F., van den Bosch, F.~C., Gallazzi, A., \& Rix, H.-W.\ 2009, \apj, 698, 1232 

\bibitem[van Dokkum et al.(2008)]{vandokkum} van Dokkum, P.~G., et al.\ 2008, \apjl, 677, L5 



\end{thebibliography}
\end{document}